\documentstyle{elsart}

\input epsf

\newcommand\lesssim{\mathrel{\rlap{\lower4pt\hbox{\hskip1pt$\sim$}}
    \raise1pt\hbox{$<$}}}
\newcommand\gtrsim{\mathrel{\rlap{\lower4pt\hbox{\hskip1pt$\sim$}}
    \raise1pt\hbox{$>$}}}

\begin{document}
\begin{frontmatter}
\title{Cosmological Constraints from Primordial Black Holes}
\author{Andrew R.~Liddle and Anne M.~Green},
\address{Astronomy Centre, University of Sussex, Brighton BN1 9QJ, UK}

\begin{abstract}
Primordial black holes may form in the early Universe, for example from the 
collapse of large amplitude density perturbations predicted in some 
inflationary models. Light black holes undergo Hawking evaporation, the 
energy injection from which is constrained both at the epoch of 
nucleosynthesis and at the present. The failure as yet to unambiguously 
detect primordial black holes places important constraints. In this article, 
we are particularly concerned with the dependence of these constraints on the 
model for the complete cosmological history, from the time of formation to 
the present. Black holes presently give the strongest constraint on the 
spectral index $n$ of density perturbations, though this constraint does 
require $n$ to be constant over a very wide range of scales.
\end{abstract}

\begin{keyword}
Black holes \sep inflationary cosmology 
\PACS 97.60.Lf \sep 98.80.Cq 
\end{keyword}
\end{frontmatter}

\section{Introduction}

Black holes are tenacious objects, and any which form in the very early 
Universe are able to survive until the present, unless their Hawking 
evaporation is important. The lifetime of an evaporating black hole is given 
by 
\begin{equation}
\frac{\tau}{10^{17} \, {\rm sec}} \simeq \left( \frac{M}{10^{15} \, {\rm
	grams}} \right)^3 \,.
\end{equation}
From this we learn that a black hole of initial mass $M \sim 10^{15}$g will  
evaporate at the present epoch, while for significantly heavier black holes 
Hawking evaporation is negligible. Another mass worthy of consideration is 
$M \sim 10^{9}$g, which leads to evaporation around the time of 
nucleosynthesis, which is well enough understood to tolerate only modest 
interference from black hole evaporation by-products. 

Several mechanisms have been proposed which might lead to black hole 
formation; the 
simplest is collapse from large-amplitude, short-wavelength density 
perturbations. They will form with approximately the horizon mass, which in a 
radiation-dominated era is given by
\begin{equation}
\label{hormass}
M_{{\rm HOR}} \simeq 10^{18} \, {\rm g} \, \left( 
	\frac{10^7 \, {\rm GeV}}{T} \right)^2 \,,
\end{equation}
where $T$ is the ambient temperature.
This tells us that any black holes for which evaporation is important must 
have formed during very early stages of the Universe's evolution. In 
particular, formation corresponds to times much earlier than nucleosynthesis 
(energy scale of abut $1\,$MeV), which is the earliest time that we have any 
secure knowledge concerning the evolution of the Universe. Any modelling of 
the evolution of the Universe before one second is speculative, and 
especially above the electro-weak symmetry breaking scale (about $100 \,$GeV) 
many possibilities exist. Note also that although we believe we understand 
the relevant physics up to the electro-weak scale, the cosmology between that 
scale and nucleosynthesis could still be modified, say by some massive but 
long-lived particle. In this article we will consider the standard cosmology  
and two alternatives \cite{gl,glr}.

We define the mass fraction of black holes as
\begin{equation}
\beta \equiv \frac{\rho_{{\rm pbh}}}{\rho_{{\rm tot}}} \,,
\end{equation}
and will use subscript `i' to denote the initial values. In fact, we will 
normally prefer to use
\begin{equation}
\alpha \equiv \frac{\rho_{{\rm pbh}}}{\rho_{{\rm tot}}-\rho_{{\rm pbh}}} =
	\frac{\beta}{1-\beta} \,,
\end{equation}
which is the ratio of the black hole energy density to the energy density of 
everything else.
Black holes typically offer very strong constraints because after formation 
the black hole energy density redshifts away as non-relativistic matter 
(apart from the extra losses through evaporation). In the standard cosmology 
the Universe is radiation dominated at these times, and so the energy density 
in black holes grows, relative to the total, proportionally to the scale 
factor $a$. As interesting black holes form so early, this factor can be 
extremely large, and so typically the initial black hole mass fraction is 
constrained to be very small. 

The constraints on evaporating black holes are well known, and we summarize 
them in Table~\ref{massfrac}. This table shows the allowed mass fractions at 
the time of evaporation. An additional, optional, constraint can be imposed 
if one imagines that black hole evaporation leaves a relic particle, as these 
relics must then not over-dominate the mass density of the present Universe 
\cite{BCL:PBH}.
For black holes massive enough to have negligible evaporation, the mass 
density constraint is the only important one (though in certain mass ranges 
there are also microlensing limits which are somewhat stronger).

\begin{table}[t]
\caption[massfrac]{\label{massfrac} Limits on the mass fraction of black 
holes at
evaporation.}
\begin{tabular}{|c|c|c|}
\hline
\hline
Constraint & Range & Reason \\ 
\hline 
$\alpha_{\rm{evap}} < 0.04$ & $10^{9}$ g $< M < 10^{13}$ g & 
Entropy per baryon\\
& & at nucleosynthesis \cite{var:ent} \\ 
\hline 
$\alpha_{\rm{evap}} < 10^{-26} \frac{M}{m_{{\rm Pl}}}$ & 
$M \simeq 5\times10^{14}$~g & $\gamma$ rays
from current\\
& & explosions \cite{var:gam} \\ 
\hline $\alpha_{\rm{evap}} < 6\times10^{-10} \left( \frac{M}{m_{{\rm 
Pl}}}\right)^{1/2}$
 & $10^{9}$~g $ < M <10^{11}$~g & n$\bar{\rm{n}}$ production \\
 & & at nucleosynthesis \cite{var:neu} \\ 
\hline
$\alpha_{\rm{evap}} < 5\times10^{-29} \left( \frac{M}{m_{{\rm 
Pl}}}\right)^{3/2}$ &
$10^{10}$~g $< M < 10^{11}$~g & Deuterium destruction \cite{lin:deu} 
\\ \hline
$\alpha_{\rm{evap}} < 1\times10^{-59}\left( \frac{M}{m_{{\rm 
Pl}}}\right)^{7/2}$ &
$10^{11}$~g $< M < 10^{13}$~g & Helium-4 spallation \cite{var:he4}\\
\hline
\end{tabular}
\vspace*{2pc}
\end{table}

We will study three different cosmological histories in this paper, all of 
which are currently observationally viable. The first, which we call the 
standard cosmology, is the minimal scenario. It begins at some early time 
with cosmological inflation, which is necessary in order to produce the 
density perturbations which will later collapse to form black holes. 
Inflation ends, through the preheating/reheating transition (which we will 
take to be short), giving way to a period of radiation domination. Radiation 
domination is essential when the Universe is one second old, in order for 
successful nucleosynthesis to proceed. Finally, radiation domination gives 
way to matter domination, at a redshift $z_{{\rm eq}} = 24\,000\,\Omega_0 
h^2$ where $\Omega_0$ and $h$ have their usual meanings, to give our present 
Universe. 

The two modified scenarios replace part of the long radiation-dominated era 
between the end of inflation and nucleosynthesis. The first possibility is 
that there is a brief second period of inflation, known as thermal inflation 
\cite{ls}. Such a period is unable to generate significant new density 
perturbations, but may be desirable in helping to alleviate some relic 
abundance problems not solved by the usual period of high-energy inflation. 
The second possibility is a period of matter-domination brought on by a 
long-lived massive particle, whose eventual decays restore radiation 
domination before nucleosynthesis. For definiteness, we shall take the 
long-lived particles to be the moduli fields of superstring theory, though 
the results apply for any non-relativistic decaying particle.

\section{The standard cosmology}

\begin{figure}[t]
\centering 
\leavevmode\epsfysize=10cm \epsfbox{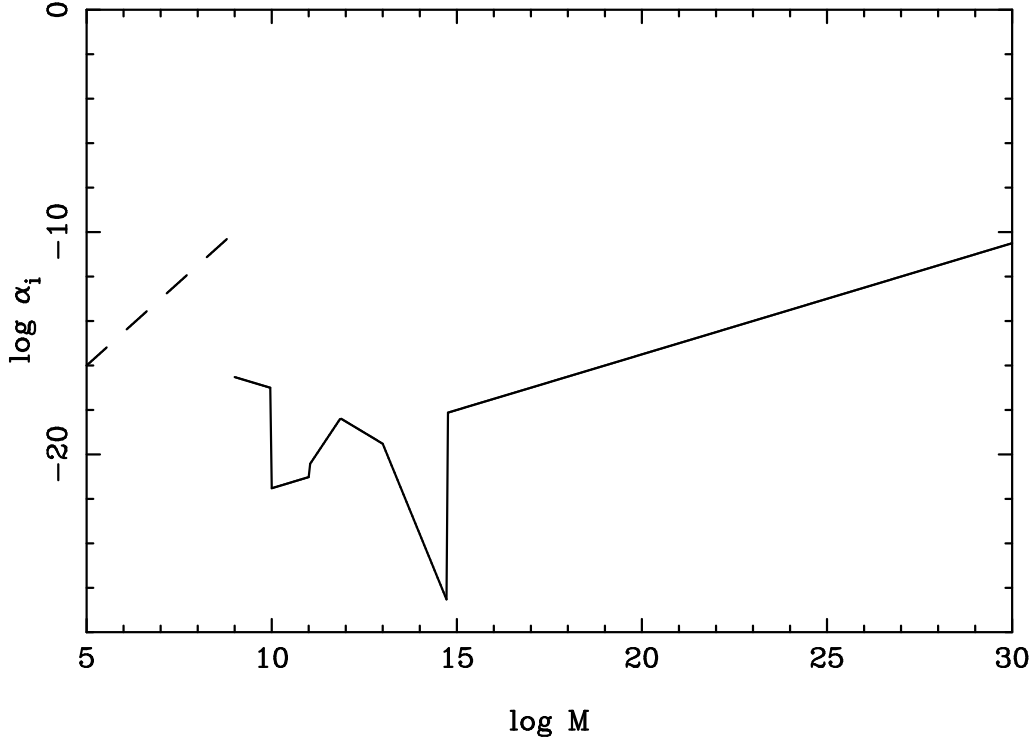}\\ 
\caption[fig1]{\label{fig1} Here $\alpha_{{\rm i}}$ is the initial fraction 
of black holes permitted to form. The dotted line assumes black hole 
evaporation leaves a Planck-mass relic, and is optional.} \vspace{2pc}
\end{figure}

Once the cosmology is fixed, the limits on the mass fraction at evaporation
shown in Table~\ref{massfrac}, along with the constraints from the present
mass density, are readily converted into limits on the initial mass fraction
\cite{limits,gl}.  The limits in different mass ranges are shown in
Figure~\ref{fig1}; typically no more than about $10^{-20}$ of the mass of the
Universe can go into black holes at any given epoch.  This limits the size of 
density perturbations on the relevant mass scale.  

These limits apply down to the lightest black hole which is able to form, 
which is governed by the horizon size at the end of inflation.

The black hole constraint limits density perturbations on scales much shorter 
than conventional measures of the density perturbation spectrum using 
large-scale structure and the microwave background (though the scales 
corresponding to the black hole formation are similar to those on which 
gravitational waves may be probed by the {\sc ligo}, {\sc virgo} and {\sc 
geo} interferometer projects \cite{cllw}). However, an interesting 
application of the black hole constraints shown in Figure~\ref{fig1} can be 
made if one has a definite form for the power spectrum. The simplest 
possibility is a power-law spectrum, whose spectral index $n$ is assumed to 
remain constant across all scales, the interesting case being where $n>1$ --- 
a so-called blue spectrum. The constancy of $n$ is in fact predicted by some 
hybrid inflation models, which are the most natural way of obtaining a blue 
spectrum.

With $n>1$ the shortest-scale perturbations dominate, and the black hole 
constraints were explored by Carr et al.~\cite{cgl}, who found that $n$ was 
limited to be less than around $1.4$ 
to $1.5$. We have redone their analysis and corrected two significant errors. 
First, they used an incorrect scaling of the horizon mass during the 
radiation era, which should read
\begin{equation}
\sigma_{{\rm hor}}(M) = \sigma_{{\rm hor}}(M_{{\rm eq}}) \left( 
	\frac{M}{M_{{\rm eq}}} \right)^{(1-n)/4} \,.
\end{equation}
Secondly, their normalization of the spectrum to the COBE observations, to 
fix the long wavelength behaviour, was incorrect (too low) by a factor of 
around twenty. With these corrections, the constraint on $n$ tightens 
considerably to become $n \lesssim 1.25$ \cite{gl}. 

Ignoring for the time being uncertainties in cosmological modelling, this 
should be regarded as a very hard limit; it is not useful to try and think of 
it as representing some confidence level. Because the density perturbations 
are assumed gaussian, the formation rate of black holes is extremely 
sensitive to the amplitude of perturbations on the scale under consideration. 
Hence a very modest change in $n$ gives a huge change in the predicted black 
hole abundance, which makes a rapid transition between totally negligible and 
enormously excessive as $n$ crosses the quoted limit. In fact, to obtain a 
black hole density near the present limits requires a considerable 
fine-tuning, as we saw in Figure~\ref{fig1} that only about $10^{-20}$ or so 
of the mass of the Universe must be channelled into black holes.

However, the previous paragraph did not take into account uncertainties in 
cosmological modelling, and that is what we will quantify in the remainder of 
this article. We will also note that a lifting of the gaussianity assumption, 
presently a controversial topic \cite{BP,nong}, makes little change.

\section{With thermal inflation}

\begin{figure}[t]
\centering 
\leavevmode\epsfysize=10cm \epsfbox{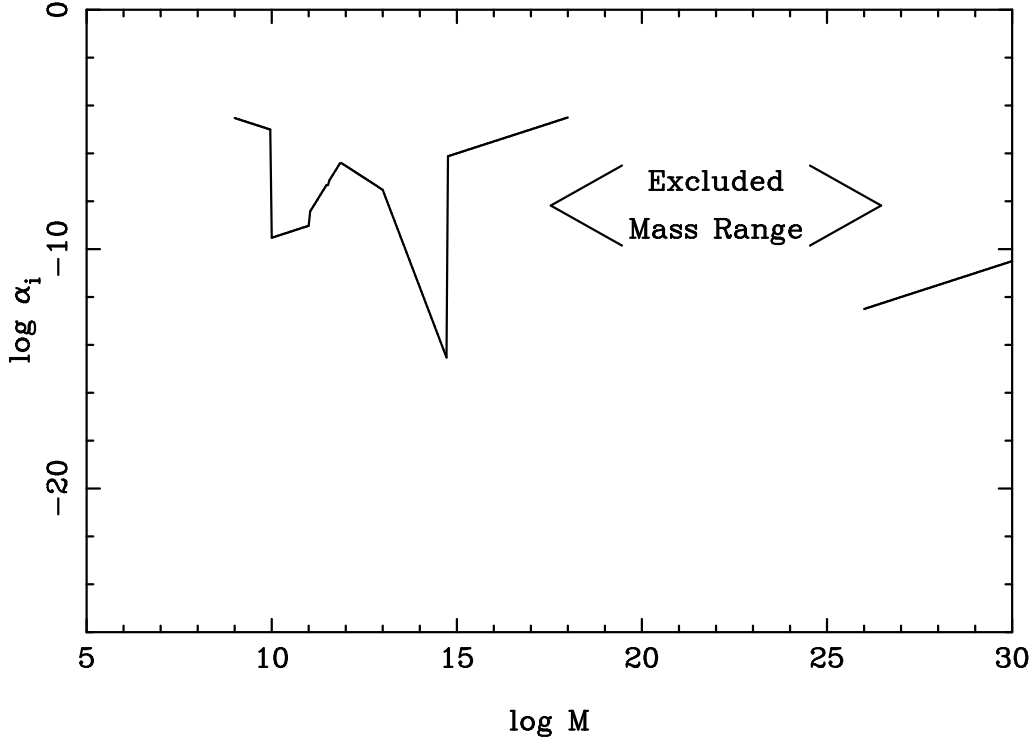}\\ 
\caption[fig2]{\label{fig2} Black hole constraints modified to include 
thermal 
inflation.} 
\vspace{1pc}
\end{figure} 

Thermal inflation is a brief period of inflation occurring at an intermediate
energy scale.  We model it as occurring from $T = 10^7$ GeV down to the
supersymmetry scale $T = 10^3$ GeV, then reheating back up to $10^7$ GeV,
which is the standard thermal inflation scenario.  It drives $\ln(10^7 \, 
{\rm GeV}/10^3 \, {\rm
GeV}) \simeq 10$ $e$-foldings of inflation.  As we have seen
[Eq.~(\ref{hormass}], most of the interesting mass region contains black 
holes
forming before $T = 10^7$ GeV, which implies that the constraints are
affected by thermal inflation.  Three effects are important:
\vspace*{-8pt}
\begin{itemize}
\item Dilution of black holes during thermal inflation.
\item A change in the correspondence of scales: COBE scales 
leave the horizon closer to the end of inflation.
\item A mass range which enters the horizon before thermal 
inflation, but leaves again during it. No new perturbations are generated on 
this scale during thermal inflation, so from the horizon mass formula we find 
a missing mass range between $10^{18}$g and $10^{26}$g in which black holes 
won't form. Thermal inflation at higher energy could exclude masses below 
$10^{15}$g.
\end{itemize}
\vspace*{-8pt}
The dilution effect is shown in Figure~\ref{fig2}; typical constraints now 
lie around $10^{-10}$ rather than $10^{-20}$. Taking all the effects into 
account, the constraint on the spectral index weakens to $n \lesssim 
1.3$ \cite{gl}.

\begin{figure}[t]
\centering 
\leavevmode\epsfysize=10cm \epsfbox{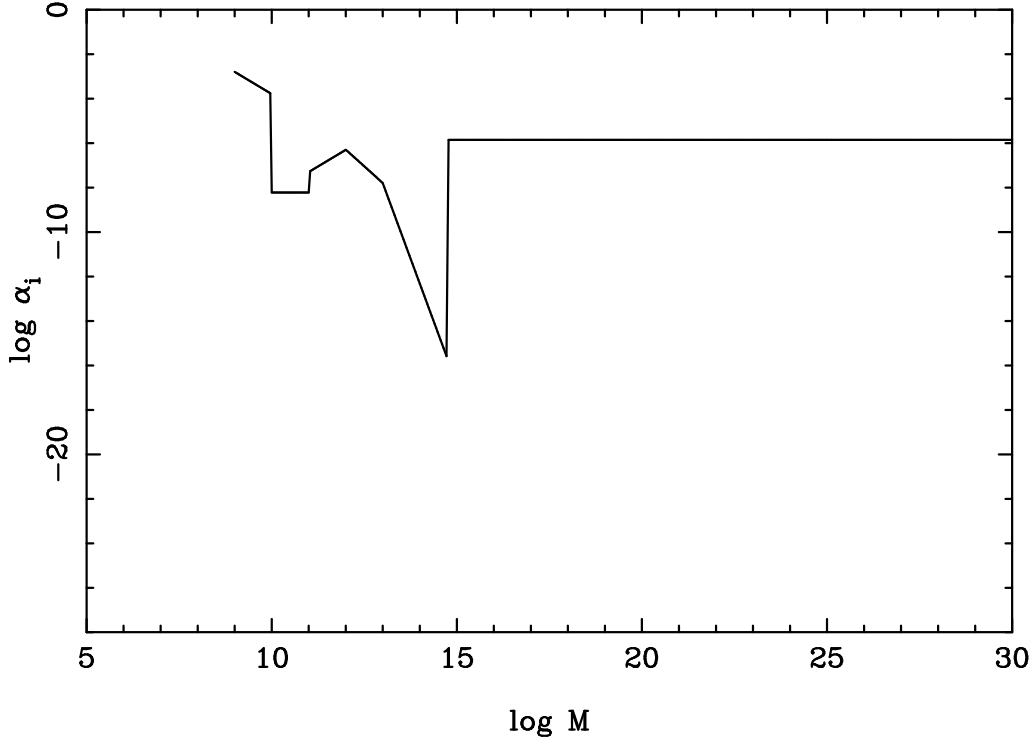}\\ 
\caption[fig3]{\label{fig3} Black hole constraints modified for prolonged 
moduli 
domination.} 
\vspace{1pc}
\end{figure} 

\section{Cosmologies with moduli domination}

A prolonged early period of matter 
domination is another possible modification to the standard cosmology 
\cite{glr}. For 
example, moduli fields may dominate, and in certain parameter regimes can 
decay before nucleosynthesis. Various assumptions are possible; here we'll 
assume moduli domination as soon as they start to oscillate (around 
$10^{11}\,$GeV). Part of the interesting range of black hole masses forms 
during moduli 
domination rather than radiation domination. Figure~\ref{fig3} shows the 
constraints in this case, and with moduli domination the limit on $n$ again 
weakens to $n \lesssim 1.3$ \cite{glr}.

\section{Conclusions}

Although black hole constraints are an established part of modern cosmology, 
they 
are sensitive to the entire cosmological evolution. In the standard 
cosmology, a power-law spectrum is constrained to $n < 1.25$, presently the 
strongest observational constraint on $n$ from any source. Alternative 
cosmological histories can weaken this to $n < 1.30$, and worst-case 
non-gaussianity \cite{BP} can weaken this by another 0.05 or so, though 
hybrid models giving constant $n$ give gaussian perturbations. Finally, we 
note that while the impact of the cosmological history on the 
density perturbation constraint is quite modest due to the exponential 
dependence of the
formation rate, the change can be much more significant for other formation 
mechanisms, such as cosmic strings where the black hole formation rate is a 
power-law of the mass per unit length $G\mu$. After all, the permitted 
initial mass density of black holes does increase by many orders of magnitude 
in these alternative cosmological models.

\section*{Acknowledgments}   
ARL was supported by the Royal Society and AMG by PPARC. We thank Toni Riotto 
for collaboration on the analysis of the mod\-uli-dominated cosmology, and 
Bernard Carr and Jim Lidsey for discussions.


\begin{thebibliography}{9}
\bibitem{gl} A. M. Green and A. R. Liddle, Phys. Rev. D {\bf 56} (1997) 6166.
\bibitem{glr} A. M. Green, A. R. Liddle and A. Riotto, Phys. Rev. D 
	{\bf 56} (1997) 7559.
\bibitem{BCL:PBH} J. D. Barrow, E. J. Copeland and A. R. Liddle,
	Phys. Rev. D {\bf 46} (1992) 645.
\bibitem{var:ent} S. Mujana and K. Sato,
         Prog. Theor. Phys. {\bf 59} (1978) 1012; B. V. Vainer and
         P. D. Nasselskii, Astron. Zh {\bf 55} (1978) 231
         [Sov. Astron. {\bf 22} (1978) 138].
\bibitem{var:gam} J. H. MacGibbon, Nature {\bf 320} (1987) 308;
         J. H. MacGibbon and B. Carr, Astrophys. J. {\bf 371} (1991) 447.
\bibitem{var:neu} Ya. B. Zel'dovich, A. A. Starobinsky, M. Y. 
         Khlopov and V. M. Chechetkin, Pis'ma Astron. Zh. {\bf 3} (1977)
         308 [Sov Astron. Lett. {\bf 22} (1977) 110].
\bibitem{lin:deu} D. Lindley, Mon. Not. R. Astron. Soc. {\bf 193} (1980)
         593.
\bibitem{var:he4} B. V. Vainer, D. V. Dryzhakova and P. D. Nasselskii,
         Pis'ma Astron. Zh. {\bf 4} (1978) 344 [Sov. Astron. Lett
        {\bf 4} (1978) 185].        
\bibitem{ls} D. H. Lyth and E. D. Stewart, Phys. Rev. Lett. {\bf 75} (1995)
	201.
\bibitem{limits} B. J. Carr, in {\em Observational and Theoretical
	Aspects of Relativistic Astrophysics and Cosmology} edited by
	J. L. Sanz and L. J. Goicoechea (World Scientific, Singapore,
	1985).
\bibitem{cllw} E. J. Copeland, A. R. Liddle, J. E. Lidsey and D. Wands,
	Sussex preprint gr-qc/9803070 (1998).
\bibitem{cgl} B. J. Carr, J. H. Gilbert and J. E. Lidsey,
	Phys. Rev. D {\bf 50} (1994) 4853.
\bibitem{BP} J. S. Bullock and J. R. Primack, Phys. Rev. D {\bf 55} (1997) 
	7423.
\bibitem{nong} P. Ivanov, TAC preprint astro-ph/9708224 (1997). 

\end{thebibliography}
\end{document}